\documentclass[twocolumn,superscriptaddress,showpacs,preprintnumbers,amsmath,amssymb]{revtex4-1}

\usepackage{graphicx}
\usepackage{dcolumn}
\usepackage{bm}
\usepackage{color}
\begin{document}

\title{Probing for massive stochastic gravitational-wave background with a detector network}

\author{Atsushi~Nishizawa}
\email{nishizawa@tap.scphys.kyoto-u.ac.jp}
\affiliation{Department of Physics, Kyoto University, Kyoto, 606-8502, Japan}
\author{Kazuhiro~Hayama}
\affiliation{Osaka City University,
Graduate School of Science, Faculty of Science
3-3-138 Sugimoto Sumiyoshi-ku Osaka-shi, 558-8585, Japan}
\affiliation{TAMA project, National Astronomical Observatory of Japan, Mitaka, Tokyo 181-8588, Japan}
\date{\today}

\begin{abstract}
In a general metric theory of gravitation in four dimensions, six polarizations of a gravitational wave are allowed: two scalar and two vector modes, in addition to two tensor modes in general relativity. Such additional polarization modes appear due to additional degrees of freedom in modified gravity theories. Also graviton mass, which could be different in each polarization, is another characteristic of modification of gravity. Thus, testing the existence of additional polarization modes and graviton mass can be a model-independent test of gravity theories. Here we extend the previous framework of correlation analysis of a gravitational-wave background to the massive case and show that a ground-based detector network can probe for massive stochastic gravitational waves with its mass around $\sim10^{-14}\,\rm{eV}$. We also show that more than three detectors can cleanly separate the mixture of polarization modes in detector outputs and determine the graviton mass.
\end{abstract}
\maketitle

\section{Introduction}
In a past decade, direct detection experiments of a gravitational wave (GW) have been well developed. The next-generation kilometer-scale laser-interferometric GW detectors such as advanced LIGO (aLIGO) \cite{aLIGO:web}, advanced VIRGO (aVIRGO) \cite{aVIRGO:web}, and KAGRA \cite{KAGRA:web} are under construction and will start observation within the coming 3-5 years, aiming at the first detection of a GW. The GW observation not only brings us valuable information about astronomy and cosmology but also allows us to test a gravity theory.

If a gravity theory deviates from general relativity (GR), GWs are also modified in (i) GW waveform, mainly phase evolution, (ii) additional polarization modes, and (iii) graviton mass or propagation speed (For review, see \cite{Broeck:2013kx,Gair:2012nm,Yunes:2013dva}). The first one is associated with dynamics of a compact binary. A gravity test in this direction has been investigated by many authors (See \cite{Yunes:2013dva} and Refs. therein). However, most works rely on the matched filtering technique \cite{Cutler:1994ys}, which requires accurate wave templates in a specific model of a gravity theory and is model-dependent method. Some exceptions are model-independent studies focusing on deviation from GR \cite{Mishra:2010tp,DelPozzo:2011pg,Li:2011cg,Yunes:2009ke,Yunes:2010qb,Cornish:2011ys}. Secondly, additional polarization modes arise due to extra degrees of freedom appearing in a modified gravity theory. In GR, a GW has two tensor polarization modes (plus and cross modes), while in a general metric theory of gravitation, the GW is allowed to have at most six polarizations, including scalar and vector modes \cite{Eardley:1973,Will:2005va}. The number of polarizations in various gravity theories has been studied in \cite{Alves:2009eg,Alves:2010ms,deRham:2011qq} and the model-independent test has been proposed in \cite{Nishizawa:2009bf,Nishizawa:2009jh,Chatziioannou:2012rf,Hayama:2012au}. The third one, graviton mass changes the propagation speed of a GW. Namely, the dispersion relation of a GW is modified and a GW undergoes mass-dependent phase evolution during its propagation. Constraint on graviton mass from the observation of compact binary coalescences has been originally proposed in \cite{Will:1997bb} and later forecasted with more accurate GW waveforms including spin-orbit and spin-spin couplings and spin precession effects \cite{Yagi:2009zm,Stavridis:2009mb} (See also the recent paper \cite{Berti:2011jz} and Refs. therein). Also mass constraint will be obtained from the observation of a GWB in pulsar timing array \cite{Lee:2010cg}. Since the polarization and massive graviton tests do not demand one to know neither dynamics of astronomical objects nor exact waveforms, we are able to perform a model-independent test of gravity.
 
Here we consider a stochastic gravitational-wave background (GWB), which is incoherent superposition of gravitational waves produced by many unresolved astronomical sources or by inflation and reheating in the early universe. The sensitivity of a GW detector pair to a GWB with scalar polarization has been investigated in \cite{Maggiore:1999wm,Gasperini:2001mw}. However, their studies assumed the existence of a single polarization (scalar) mode. Thus it is not manifest how the mixing with other polarization modes (tensor or vector) affect the detector sensitivity to the scalar mode and the mode separability. Previously, we have studied the detection and separation of scalar, vector, and tensor polarization modes of a GWB, using a network of ground-based laser interferometers \cite{Nishizawa:2009bf} (With the pulsar timing array, see Ref. ~\cite{Lee:2008ApJ}). We found that with the correlation signals obtained from more than three detectors, the scalar, vector, and tensor modes of a GWB can be cleanly separated and detected without degeneracy between polarizations. The mode separation has been extensively studied in \cite{Nishizawa:2009jh}, varying detector relative distance and orientation and the number of detectors. Then we found general conditions required for successful mode separation and that a current network of ground-based detectors is accidentally nearly optimal. However, the previous studies assumed that graviton is massless. It is likely that graviton has mass if gravity is modified and a GW have additional polarization mode. In fact, for instance, tensor graviton is massless and scalar graviton is massive in $f(R)$ gravity \cite{Sotiriou:2008rp,DeFelice:2010aj} and scalar-tensor theory \cite{Fujii-Maeda:book}. In this paper, we further extend our previous formalism to a massive GWB containing scalar, vector, and tensor polarizations and investigate what range of graviton mass is searched by ground-based detectors and how accurately the graviton mass is determined. 

The organization of this paper is as follows. In Sec.~\ref{sec2}, we briefly review the formulation of a GWB with scalar, vector, and tensor polarizations and extend it to the massive case. We also mention current constraints on graviton mass. In Sec.~\ref{sec3}, the cross-correlation analysis of a GWB is extended to the massive case, focusing on how the standard analysis method in the massless case is altered. Sec.~\ref{sec4} is the main part of this paper and describes parameter estimation accuracy based on the Fisher information matrix. In Sec.~\ref{sec5}, we discuss the parameter estimation accuracy in more practical situation, that is, in a mixture of three polarization modes of a GWB. We devote the last section \ref{sec6} to the summary of this paper.

\section{GW in modified gravity theories}
\label{sec2}
\subsection{GW polarization modes}
In general, a metric gravity theory in four dimensions allows at most six polarization modes of a GW \cite{Eardley:1973,Will:book}. Let us define a wave orthonormal coordinate that are constructed by a unit vector $\hat{\Omega}$ directed to the propagation direction of a GW and two unit vectors $\hat{e}_{\theta}$ and $\hat{e}_{\phi}$ orthogonal to $\hat{\Omega}$ and each other. With these vectors, the polarization modes are defined as 
\begin{align}
\mathbf{e}^{+} &\equiv \hat{e}_{\theta} \otimes \hat{e}_{\theta} -\hat{e}_{\phi} \otimes \hat{e}_{\phi} \;,\\
\mathbf{e}^{\times}&\equiv \hat{e}_{\theta} \otimes \hat{e}_{\phi} +\hat{e}_{\phi} \otimes \hat{e}_{\theta} \;,\\
\mathbf{e}^{b}&\equiv \hat{e}_{\theta} \otimes \hat{e}_{\theta} +\hat{e}_{\phi} \otimes \hat{e}_{\phi} \;,\\
\mathbf{e}^{\ell}&\equiv \sqrt{2}\, \hat{\Omega} \otimes \hat{\Omega}\;,\\
\mathbf{e}^{x}&\equiv \hat{e}_{\theta} \otimes \hat{\Omega}+\hat{\Omega} \otimes \hat{e}_{\theta} \;,\\
\mathbf{e}^{y}&\equiv \hat{e}_{\phi} \otimes \hat{\Omega}+\hat{\Omega} \otimes \hat{e}_{\phi} \;,
\end{align}
where the symbol $\otimes$ denotes a tensor product. The $+$, $\times$, $b$, $\ell$, $x$, and $y$ polarization modes are called plus, cross, breathing, longitudinal, vector-x, and vector-y y $y$  modes, respectively. Each polarization mode is orthogonal to one another and is normalized so that $e_{ij}^{A}e^{ij}_{A^{\prime}}=2 \delta _{AA^{\prime}}, \; A,A^{\prime}=+,\times,b,\ell,x,y$. According to rotation symmetry about the propagation axis of a GW, the $+$ and $\times$ modes are identified with tensor-type (spin-2) GWs, the $x$ and $y$ modes are vector-type (spin-1) GWs, and the $b$ and $\ell$ modes are scalar-type (spin-0) GWs. Note that the breathing and longitudinal modes are not traceless, in contrast to the ordinary plus and cross polarization modes in GR. 

A GW with the six polarizations is expressed as 
\begin{equation}
h_{ij}(t, \vec{X}, \hat{\Omega})= \sum_A h_{A} (t,\vec{X}) e_{ij}^{A} (\hat{\Omega})\;,
\end{equation} 
where $A=+,\, \times,\, b,\, \ell,\, x,\, y$. The antenna pattern functions for each polarization mode \cite{Tobar:1999,Nishizawa:2009bf} are defined as 
\begin{align}
F_A (\hat{\Omega}) &\equiv D^{ij} e_{ij}^A
(\hat{\Omega})\:, 
\label{eq2} \\
\mathbf{D} &\equiv \frac{1}{2}\left[ \hat{u} \otimes \hat{u}- \hat{v}\otimes \hat{v}\right]\:,
\nonumber
\end{align}
where the unit vectors $\hat{u}$ and $\hat{v}$ are along two arms of a laser-interferometric detector, and $\mathbf{D}$ is a so-called detector tensor, which describes the response of the laser-interferometric detector to the polarization tensors. With these antenna pattern functions, a GW signal from a detector is written as
\begin{equation}
h(t, \vec{X},\hat{\Omega})= \sum_A h_{A} (t,\vec{X}) F_{A} (\hat{\Omega})\;.
\end{equation} 
For the derivation and the explicit expressions of the antenna pattern functions, see \cite{Nishizawa:2009bf}.


\subsection{Massive gravitational waves}
Another characteristic of a GW in modified gravity is graviton mass. If graviton possesses the mass $m_g$, its dispersion relation is altered as
\begin{equation}
\omega^2=m_g^2 +k^2 \;, \nonumber
\end{equation}
where $\omega$ is the angular frequency and $k$ is the wave number. Consequently, gravitons propagate with the group velocity less than the speed of light,
\begin{equation}
v_g(\omega;m_g) \equiv \frac{d\omega}{dk} = \sqrt{1-\frac{m_g^2}{\omega^2}} \;, \nonumber
\end{equation}
and the arrival time of the GW from a point source at each detector is delayed. On the other hand, the phase velocity is larger than the speed of light:
\begin{equation}
v_p(\omega;m_g) \equiv \frac{\omega}{k} = \left(\sqrt{1-\frac{m_g^2}{\omega^2}}\right)^{-1} \;. 
\label{eq59}
\end{equation}
Then the plane wave solution of a GW is modified as
\begin{equation}
e^{i (\omega t -\vec{k}\cdot \vec{X})} = e^{i \omega [ t -\hat{\Omega}\cdot \vec{X}/v_p(\omega;m_g)]} \;,
\label{eq50}
\end{equation} 
replacing the speed of light $c$ in GR with $v_p$. One may regard the change of the phase velocity from $c$ as the change of the effective detector distance from the origin, $\vec{x}/c \rightarrow \vec{x}/v_p$. Since $c < v_p$, the effective detector distance for a massive GW is smaller. 

For a stochastic GW background, the modification of the group velocity is not relevant but the phase velocity plays an important role. The stochastic GW signal observed at a position $\vec{x}$ at time $t$ is written as
\begin{align}
h(t,\vec{X}) &= \sum_A \int _{S^2} d\hat{\Omega} \int_{-\infty}^{\infty}df \nonumber \\
& \times \tilde{h}_A (f, \hat{\Omega})\,e^{2\pi if (t-\hat{\Omega} \cdot \vec
{X}/v_p)} F_A(\hat{\Omega})\:,  \label{eq7}
\end{align}
where $\tilde{h}_A (f, \hat{\Omega})$ is the Fourier transform of the GW amplitude in each polarization mode and the frequency is $f=\omega/(2\pi)$. The Fourier transform of Eq.\,(\ref{eq7}) is given by
\begin{equation}
\tilde{h}(f,\vec{x})=\sum_A \int_{S^2}d\hat{\Omega}\, \tilde{h}_A(f,\hat{\Omega} )e^{-2\pi if \hat{\Omega} \cdot \vec{X}/v_p} F_A (\hat{\Omega})\:.
\label{eq11}
\end{equation}

Another consequence of the dispersion relation is the existence of the lowest angular frequency of a GW:
\begin{align}
\omega_{\rm{min}} &= m_g \nonumber \\
&\approx 6.58 \times 10^{-14} \left( \frac{f_g}{100\,{\rm{Hz}}} \right) \,{\rm{eV}}\:,
\label{eq25}
\end{align}
where $f_g$ is the frequency corresponding to $m_g$. This is because below the frequency $\omega=m_g$, the wave number in the dispersion relation becomes imaginary and GWs become unstable. The reason is also understood directly from the definitions of graviton phase velocity in Eq.~(\ref{eq59}) and the plane wave solution in Eq.~(\ref{eq50}). As a result, a GW spectrum has a steep cutoff at low frequency side, independent of how the GW is generated.

\subsection{Stochastic GW spectrum}

In this paper, we assume that a stochastic GWB is (i) isotropic, (ii) independently polarized, (iii) stationary, and (iv) Gaussian, as discussed in details in \cite{Allen:1997ad}. In this case all the statistical properties of the GWB are characterized by
\begin{equation}
\langle \tilde{h}_A^{\ast} (f,\hat{\Omega} ) \tilde{h}_{A^{\prime}} (f^{\prime},\hat{\Omega^
{\prime}} ) \rangle = \delta(f-f^{\prime})\frac{1}{4\pi} \delta^2 ( \hat{\Omega},\hat
{\Omega^{\prime}}) \delta_{AA^\prime} \frac{1}{2} S_h^A (|f|)\;,
\label{eq20}
\end{equation}
where $\delta^2 ( \hat{\Omega},\hat{\Omega^{\prime}}) \equiv \delta(\phi-\phi^{\prime}) 
\delta (\cos\theta - \cos\theta^\prime )$, and $\langle \cdots \rangle$ denotes ensemble average. $S_h^A(f)$ is the one-sided power spectral density of each polarization mode. 

Conventionally, the amplitude of GWB for each polarization is characterized by an energy density per logarithmic frequency bin, normalized by the critical energy density of the Universe:
\begin{equation}
\Omega_{\rm{gw}}^A (f) \equiv \frac{1}{\rho_c}\frac{d\rho_{\rm{gw}}^A(f)}{d \ln f}\:,
\label{eq21}
\end{equation}
where $\rho_c = 3 H_0^2/8\pi G$ and the Hubble constant is $H_0=100\,h_{0}\,\rm{km\, s^{-1}\,Mpc^{-1} }$. $\Omega_{\rm{gw}}(f)$ is related to $S_h(f)$ by \cite{Allen:1997ad,Maggiore:1999vm}
\begin{equation}
\Omega_{\rm{gw}}^A (f) = \left( \frac{2 \pi^2}{3H_0^2} \right) f^3 S_h^A (f)\:.
\label{eq22}
\end{equation}
Note that the above definition is different from that in the literature \cite{Allen:1997ad,Maggiore:1999vm} by a factor of 2 since it is defined for each polarization. 

We assume that $+$ and $\times$ modes are not polarized (The detectability of circular polarizations in the polarized case has been discussed in \cite{Seto:2006hf,Seto:2006dz,Seto:2007tn,Seto:2008sr}). We also assume that $x$ and $y$ modes are not polarized. In most of the cosmological scenarios, these assumptions are valid. Then the GWB energy density of tensor, vector, and scalar modes can be written as
\begin{eqnarray}
\Omega_{\rm{gw}}^T &\equiv & \Omega_{\rm{gw}}^{+} + \Omega_{\rm{gw}}^{\times}  \quad \quad  (\Omega_{\rm{gw}}^{+} = \Omega_{\rm{gw}}^{\times} )\;, 
\label{eq23} \\
\Omega_{\rm{gw}}^V & \equiv& \Omega_{\rm{gw}}^x + \Omega_{\rm{gw}}^y \quad \quad  (\Omega_{\rm{gw}}^{x} = \Omega_{\rm{gw}}^{y} )\;, \\
\Omega_{\rm{gw}}^S &\equiv & \frac{1}{3} \biggl( \frac{1+2\kappa}{1+\kappa} \biggr) (\Omega_{\rm{gw}}^b + \Omega_{\rm{gw}}^{\ell}) \;, 
\label{eq24}
\end{eqnarray}
where the ratio of the energy density in the longitudinal mode to that in the breathing mode is characterized by the parameter $\kappa \equiv \Omega_{\rm{gw}}^{\ell}/\Omega_{\rm{gw}}^{b}$. However, we cannot determine $\kappa$ with a GW observation because the antenna pattern functions for both modes scale just by a factor of $\sqrt{2}$ and are completely degenerated. If an model of gravity theory is specified, $\kappa$ can be calculated theoretically and one can know the energy densities in the breathing and longitudinal modes from GW observation. Our definition of $\Omega_{\rm{gw}}^S$ is just for simplification of the formulation below. One should keep in mind that $\Omega_{\rm{gw}}^S$ does not exactly correspond to the physical energy density in the scalar mode but differs from the actual value up to a factor of three, depending on the ratio $\kappa$.

In this paper, we adopt the following parameterization for the spectral shape of a GWB:
\begin{equation}
\Omega_{\rm{gw}}^A (f)=\Omega_{{\rm{gw}},0}^A s(f) \left( \frac{f}{f_0} \right)^{n_t^A} \Theta [f-f_g^A] \;,
\label{eq54}
\end{equation}
where $\Theta[\cdot]$ is a step function. The shape function $s(f)$ strongly depends on concrete generation mechanism of the GWB and the later expansion history of the universe. In the standard cosmology with massless tensor gravitons, the GWB spectrum produced in slow-roll inflation has a nearly flat spectrum \cite{Turner:1993vb}. On the contrary, in a massive gravity theory, a GWB spectrum from de-Sitter inflation has a extremely sharp peak at $f=f_g^A$ when plotted as a function of frequency \cite{Gumrukcuoglu:2012wt}. If detected, one could determine graviton mass very accurately from the peak frequency. However, its detectability strongly depends on frequency resolution of a detector. According to the estimation with realistic observation time in \cite{Gumrukcuoglu:2012wt}, the peak would be smoothed in the frequency band of ground -based GW detectors, $\sim 100\,{\rm{Hz}}$, and difficult to be detected. Our purpose in this paper is not constraining a specific gravity theory but forecasting the detectability of graviton mass with a detector network as general as possible. For this reason, we assume $s(f)=1$ for simplicity hereafter. In this case, unknown parameters are the amplitude of the spectrum $\Omega_{{\rm{gw}},0}^A$, spectral index $n_t^A$, and mass of a graviton $m_g^A$ or equivalently corresponding frequency $f_g^A$ of each polarization mode. Since three polarization modes are in general independent of each other, we have 9 parameters in total.

\subsection{Constraints on graviton mass}
\label{sec:graviton-mass-constraints}
Currently graviton mass has been constrained by several observations of the galaxy \cite{Goldhaber:1974}, the solar system \cite{Talmadge:1988}, and binary pulsars \cite{Finn:2001qi} (For summary, see \cite{Berti:2011jz} and references therein). These limits are so tight that we can set graviton mass to zero in searching for stochastic GWs in the observational frequency band of ground-based detectors around $\sim100\,{\rm{Hz}}$. In other words, there is no low frequency cutoff of a GW spectrum and no difference from a massless case. However, the constraints from the galaxy and the solar system have been obtained from the observations in static gravitational fields and cannot be applied directly to GWs. While only the mass limit from binary pulsars comes from dynamical gravitational fields. According to the work by Finn and Sutton \cite{Finn:2001qi}, the limit on the mass of gravitons is $m_g < 7.6\times 10^{-20}\,{\rm{eV}}$. From Eq.~(\ref{eq25}), this dynamical bound is far below the observational frequency of ground-based detectors. However, the bound can be applied to a tensor mode and not to scalar and vector modes. If there are scalar and vector degrees of freedom in a GW sector and their masses are so light that GWs are emitted by the binary pulsar, they would extract additional energy from the binary pulsars and the orbit decays faster than predicted in GR. The absence of such an observation means that their possible mass have to be large enough to suppress gravitational radiation if the additional modes exist. Therefore graviton masses of scalar and vector GWs could be large enough to be detected with advanced detectors on the Earth without contradicting with current observations.

On the other hand, if the suppression mechanism of the fifth force such as chameleon mechanism \cite{Khoury:2003rn,Khoury:2003aq} works in order to elude the observational constraint from the solar system \cite{Bertotti:2003rm}, scalar and vector gravitons could be much more massive beyond the reach of ground-based detectors in the high-density environment of matter. Also another suppression mechanism called the Vainstein mechanism \cite{Vainshtein:1972sx,Babichev:2013usa} might conceal the deviation from GR. Additional degrees of freedom in gravity would alter the strength of gravity at cosmological scales and the growth of large scale structure. This quasi-static perturbative regime of gravity has been tested with current cosmological observations and will be tested in the future with GWs \cite{Yagi:2011bt,Camera:2013xfa}. However, no experiment has ever confirmed whether such suppression mechanisms work well in the perturbative and dynamical regime of gravity, particularly in GW itself. Thus it would be worth searching for graviton mass experimentally with GW detectors in order to directly test the suppression mechanisms of the fifth force.

\section{Cross correlation analysis}
\label{sec3}
We focus on a stochastic GWB and discuss cross correlation between a pair of detectors. To distinguish the GWB signal from stochastic detector noise independent in each detector, one has to correlate signals between two detectors. The correlation analysis has been well developed by several authors \cite{Christensen:1992wi,Flanagan:1993ix,Allen:1997ad}. In this section, we will extend the formulation to massive GWs with tensor, vector, and scalar modes.

\subsection{Formulation}
Let us consider the outputs of a detector, $s(t)=h(t)+n(t)$, where $h(t)$ and $n(t)$ are the GW signal and the noise of a detector. We assume that the amplitude of GWB is much smaller than detector noise. Cross-correlation signal $Y$ between two detectors is defined as
\begin{equation}
Y \equiv \int_{-T/2}^{T/2} dt \int_{-T/2}^{T/2} dt^{\prime}\, s_I(t) s_J(t^{\prime} ) Q(t-t^
{\prime})\;,
\label{eq8}
\end{equation}
where $s_I$ and $s_J$ are outputs from the $I$-th and $J$-th detectors, and $T$ is observation time. $Q(t-t^{\prime})$ is a filter function, which is later introduced so that signal-to-noise ratio (SNR) is maximized. In the absence of intrinsic noise correlation between detectors, the ensemble average of the correlation signal $\mu \equiv \langle Y \rangle$ has contribution only from GW signals. The derivation for massless GWs in a tensor polarization mode has been given in \cite{Allen:1997ad}. The formulation has been extended to massless GWs with tensor, vector, and scalar modes in \cite{Nishizawa:2009bf}. Further extension to massive gravitons with the three polarization modes is straightforward since the difference is merely replacing $c$ in the massless case with $v_p(f)$ as in Eq.~(\ref{eq11}).
Tracing the standard procedure of the derivation, we obtain the following form 
\begin{equation}
\mu = \frac{3H_0^2}{20\pi^2} T \sum_A \int_{-\infty}^{\infty} df |f|^{-3} \gamma_{IJ}^A(f) \Omega_{\rm{gw}}^A(f) \tilde{Q}(f) \;. 
\label{eq30}
\end{equation}
In the above equation, we defined overlap reduction functions 
\begin{align}
\gamma_{IJ}^{T}(f;m_g^T) &\equiv \frac{5}{2} \int_{S^2} \frac{d\hat{\Omega}}{4 \pi}\, (F_I^{+} F_J^{+} +F_I^{\times} F_J^{\times}) \nonumber \\
&\times \exp \left[i\,\frac{2\pi f \,\hat{\Omega}\cdot \Delta \vec{X}}{v_p(f;m_g^T)} \right] \;, \label{eq13} \\
\gamma_{IJ}^{V}(f;m_g^V) &\equiv \frac{5}{2} \int_{S^2} \frac{d\hat{\Omega}}{4 \pi}\, (F_I^{x} F_J^{x} +F_I^{y} F_J^{y}) \nonumber \\
&\times \exp \left[i\,\frac{2\pi f \,\hat{\Omega}\cdot \Delta \vec{X}}{v_p(f;m_g^V)} \right] \;, \label{eq28} \\
\gamma_{IJ}^{S}(f;m_g^S) &\equiv \frac{15}{1+2\kappa} \int_{S^2} \frac{d\hat{\Omega}}{4 \pi}\, (F_I^{b} F_J^{b} +\kappa F_I^{\ell} F_J^{\ell}) \nonumber \\
&\times \exp \left[i\,\frac{2\pi f \,\hat{\Omega}\cdot \Delta \vec{X}}{v_p(f;m_g^S)} \right] \;, \label{eq29} \\
\end{align}
where $\Delta \vec{X} \equiv \vec{X}_I-\vec{X}_J$. The overlap reduction functions represent how much degree of correlation between detectors in the GW signal is preserved and are normalized so that they give unity in the low-frequency limit for a coaligned detector pair.

The variance of the correlation signal is calculated by assuming that noises in two detectors do not correlate at all and that the magnitude of the GW signal is much smaller than that of noise. Again following the standard procedure in \cite{Allen:1997ad}, we obtain the variance of the correlation signal
\begin{align}
\sigma ^2 &\equiv  \langle Y^2 \rangle - \langle Y \rangle ^2 \approx \langle Y^2 \rangle \nonumber \\
&= \frac{T}{4} \int_{-\infty}^{\infty} df \,P_I(|f|) P_J(|f|)\, | \tilde{Q}(f) | ^2\:, \label{eq15}
\end{align}
where the one-sided power spectrum density of noise is defined by
 \begin{equation}
\langle \tilde{n}^{\ast}_I(f) \tilde{n}_I(f^{\prime}) \rangle \equiv \frac{1}{2}\delta (f-f^
{\prime})P_I(|f|)\;. \nonumber 
\end{equation}

Now we can determine the form of the optimal filter $\tilde{Q}(f)$ and derive the SNR formula. Equations (\ref{eq30}) 
and (\ref{eq15}) are expressed more simply, using an inner product
\begin{equation}
(A, B) \equiv \int_{-\infty}^{\infty} df A^{\ast}(f) B(f) P_I(|f|) P_J(|f|)\:,
\nonumber
\end{equation} 
as
\begin{eqnarray}
\mu &=& \frac{3H_0^2}{20\pi^2}T \left( \tilde{Q}, \sum_A \frac{\gamma_{IJ}^A (|f|) \Omega_{\rm{gw}}^A(|f|)}{|
f|^3 P_I(|f|) P_J(|f|)} \right)\:,
\label{eq16} \\
\sigma ^2 &\approx & \frac{T}{4} \left( \tilde{Q}, \tilde{Q} \right)\:,
\label{eq17} 
\end{eqnarray}
From Eqs.\,(\ref{eq16}) and (\ref{eq17}), the SNR for GWB is defined as
${\rm{SNR}} \equiv \mu / \sigma $. Therefore, the optimal filter function and the SNR turn out to be
\begin{equation}
{\rm{SNR}} = \frac{3H_0^2}{10\pi^2 } \sqrt{T} \left[  \int_{-\infty}^{\infty} df \frac{\{ \sum_A \gamma_{IJ}^A (|f|) 
\Omega^A_{\rm{gw}}(|f|) \}^2}{f^6 P_I(|f|) P_J(|f|)} \right]
^{1/2}\;,
\label{eq18}
\end{equation}
and
\begin{equation}
\tilde{Q}(f)=K \, \sum_A \frac{\gamma_{IJ}^A (f) \Omega_{\rm{gw}}^A (|f|)}{|f|^3 P_I(|f|) P_J(|f|)},
\label{eq19}
\end{equation} 
with an arbitrary normalization factor $K$.

\subsection{Overlap reduction function}
The overlap reduction functions in Eqs.\,(\ref{eq13}) - (\ref{eq29}) can be written in a more convenient form by performing the angular integrals after expanding with tensorial bases. These expressions are basically the same as those in the massless case except for the replacement of $c$ with $v_p$. However, we summarize the expressions here for later use of them.

\begin{figure}[t]
\begin{center}
\includegraphics[width=6cm]{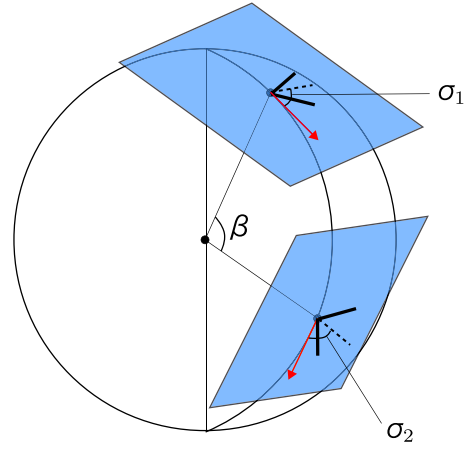}
\caption{Coordinate system on the Earth for a detector pair.}
\label{fig5}
\end{center}
\end{figure}

We introduce the coordinate system on the Earth shown in Fig.\,\ref{fig5}. The relative location and orientation of two detectors are characterized by the three parameters, $(\beta,\;\sigma_1,\;\sigma_2)$. The $\beta$ is the separation angle between two detectors, measured from the center of the Earth. The angles $\sigma_1$ and $\sigma_2$, are the orientations of the bisector of two arms of each detector, measured in a counterclockwise manner relative to the great circle connecting the two detectors. The distance between two detectors is
\begin{equation}
|\Delta \mathbf{X}| = 2 R_E \,\sin\frac{\beta}{2}\;, \nonumber
\end{equation}
where $R_E$ is the radius of the Earth and we use $R_E= 6371 \,\rm{km}$. Defining new parameters, $\sigma_{+}\equiv (\sigma_1+\sigma_2)/2$ and $\sigma_{-}\equiv (\sigma_1-\sigma_2)/2$, and 
\begin{equation}
\alpha(f) \equiv \frac{2 \pi f |\Delta \vec{X}|}{v_p(f;m_g)}\;,
\end{equation}
instead of frequency $f$, the overlap reduction functions in Eqs.\,(\ref{eq13}) - (\ref{eq29}) can be reduced in this coordinate system to
\begin{itemize}
\item{Tensor mode}\\
\begin{align}
\gamma^T (\alpha,\, \beta,\,\sigma_{+},\,\sigma_{-} ) &= \Theta _{T+}(\alpha,\, \beta)\,\cos(4\sigma_{+}) \nonumber \\
&+ \Theta _{T-}(\alpha,\, \beta)\,\cos(4\sigma_{-}) \;,
\label{eq38}
\end{align}
\begin{eqnarray}
\Theta _{T+} (\alpha,\,\beta) &\equiv & - \left( \frac{3}{8} j_0 -\frac{45}{56} j_2 + \frac{169}{896} j_4 \right) \nonumber \\
&&+ \left( \frac{1}{2} j_0 -\frac{5}{7} j_2 - \frac{27}{224} j_4 \right) \cos \beta \nonumber \\ 
&&- \left( \frac{1}{8} j_0 +\frac{5}{56} j_2 + \frac{3}{896} j_4 \right) \cos 2\beta \;, \label{eq32} \\
\Theta _{T-} (\alpha,\,\beta) &\equiv & \left( j_0 +\frac{5}{7} j_2 + \frac{3}{112} j_4 \right) \cos^4 \left( \frac{\beta}{2} \right) \;,
\label{eq33}
\end{eqnarray}
\item{Vector mode}\\
\begin{align}
\gamma^V (\alpha,\,\beta,\,\sigma_{+},\,\sigma_{-})&= \Theta _{V+}(\alpha,\,\beta)\,\cos(4\sigma_{+}) \nonumber \\
&+ \Theta _{V-}(\alpha,\,\beta)\,\cos(4\sigma_{-}) \;, 
\label{eq39}
\end{align}
\begin{eqnarray}
\Theta _{V+} (\alpha,\,\beta) &\equiv & - \left( \frac{3}{8} j_0 +\frac{45}{112} j_2 - \frac{169}{224} j_4 \right) \nonumber \\
&&+ \left( \frac{1}{2} j_0 +\frac{5}{14} j_2 + \frac{27}{56} j_4 \right) \cos \beta \nonumber \\ 
&&- \left( \frac{1}{8} j_0 -\frac{5}{112} j_2 - \frac{3}{224} j_4 \right) \cos 2\beta \;,  \label{eq34} \\
\Theta _{V-} (\alpha,\,\beta) &\equiv & \left( j_0 -\frac{5}{14} j_2 - \frac{3}{28} j_4 \right) \cos^4 \left( \frac{\beta}{2} \right) \;,
\label{eq35}
\end{eqnarray}
\item{Scalar mode}\\
\begin{align}
\gamma^S (\alpha,\,\beta,\,\sigma_{+},\,\sigma_{-})&= \Theta _{S+}(\alpha,\,\beta)\,\cos(4\sigma_{+}) \nonumber \\
&+ \Theta _{S-}(\alpha,\,\beta)\,\cos(4\sigma_{-}) \;, 
\label{eq40}
\end{align}
\begin{eqnarray}
\Theta _{S+} (\alpha,\,\beta) &\equiv & - \left( \frac{3}{8} j_0 +\frac{45}{56} j_2 + \frac{507}{448} j_4 \right) \nonumber \\
&&+ \left( \frac{1}{2} j_0 +\frac{5}{7} j_2 - \frac{81}{112} j_4 \right) \cos \beta \nonumber \\ 
&&- \left( \frac{1}{8} j_0 -\frac{5}{56} j_2 + \frac{9}{448} j_4 \right) \cos 2\beta \;, \label{eq36} \\
\Theta _{S-} (\alpha,\,\beta) &\equiv & \left( j_0 -\frac{5}{7} j_2 + \frac{9}{56} j_4 \right) \cos^4 \left( \frac{\beta}{2} \right) \;,
\label{eq37}
\end{eqnarray}
\end{itemize}
where $j_n (\alpha)$ is the spherical Bessel function with the argument $\alpha$. 

\begin{table}[t]
\begin{center}
\begin{tabular}{|l|c|c|c|c|c|}
\hline
detector pair & $\beta$ & $\sigma_{+}$ & $\sigma_{-}$ & separation [km] & $f_c$ [Hz] \\
\hline
K - H & 72.4 & 25.6 & 89.1 & 7.52 $\times 10^3$ & 6.3 \\
K - L & 99.2 & 68.1 & 42.4 & 9.71 $\times 10^3$ & 4.8 \\
K - V & 86.6 & 5.6 & 28.9 & 8.74 $\times 10^3$ & 5.4 \\
H - L & 27.2 & 62.2 & 45.3 & 3.00 $\times 10^3$ & 16 \\
H - V & 79.6 & 55.1 & 61.1 & 8.16 $\times 10^3$ & 5.7 \\
L - V & 76.8 & 83.1 & 26.7 & 7.91 $\times 10^3$ & 6.0 \\
\hline
\end{tabular}
\end{center}
\caption{Relative positions and orientations of detector pairs on the Earth (in unit of degree), and separation between two detectors and the characteristic frequency of the overlap reduction function.}
\label{tab2}
\end{table}

\begin{figure*}[t]
\begin{center}
\includegraphics[width=16cm]{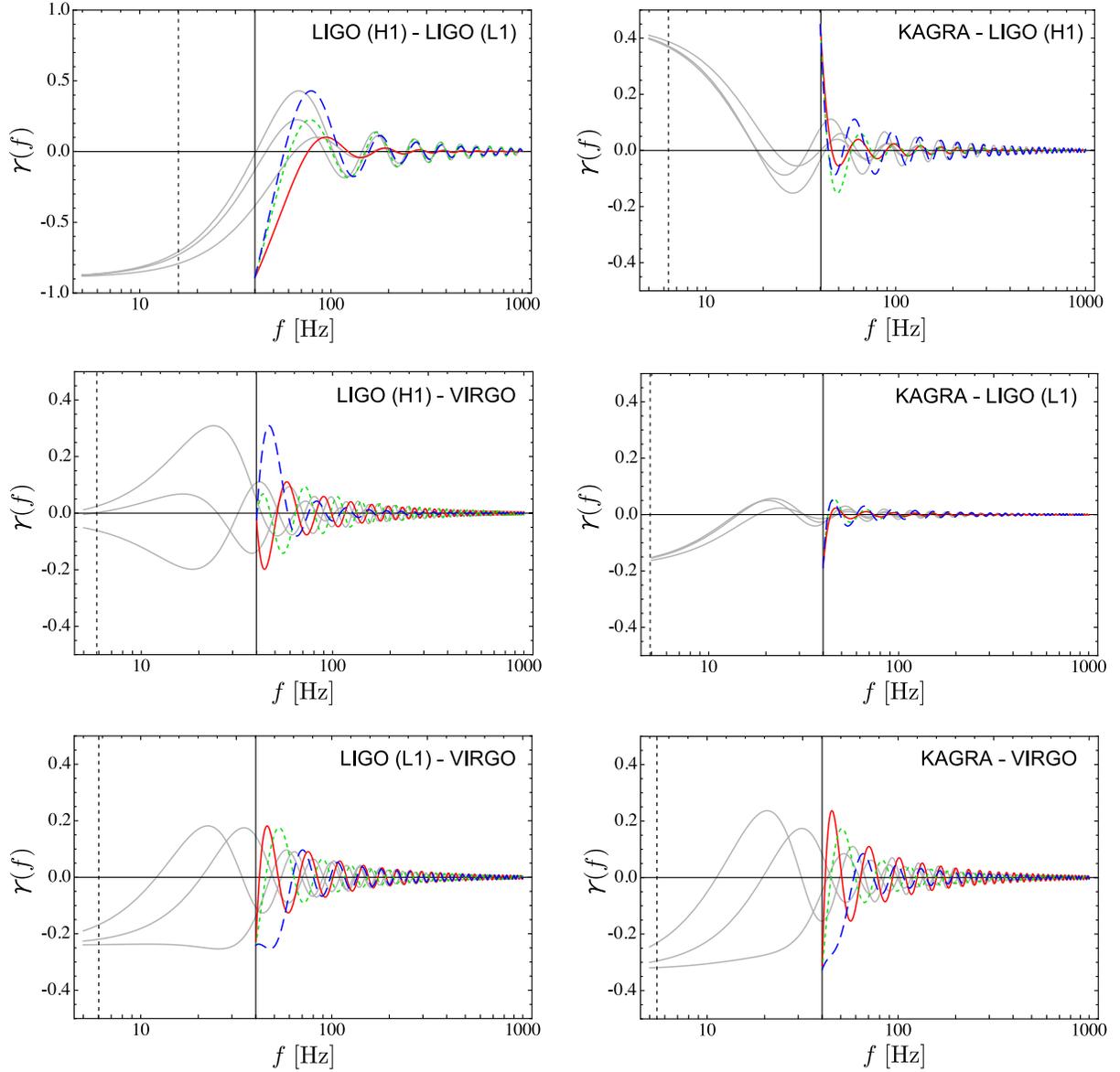}
\caption{Overlap reduction functions for real-detector pairs on the Earth. The frequency corresponding to graviton mass is set to $f_g=40\,{\rm{Hz}}$ ($m_g\approx 2\times 10^{-14}\,{\rm{eV}}$) (vertical solid line) for illustration. The characteristic frequency $f_c$, which is relevant for a massless overlap reduction function, is also shown (vertical dashed line). Each curve shows tensor mode (red, solid), vector mode (green, dotted), and scalar mode (blue, dashed). The overlap reduction functions for massless graviton case are also shown with monochromatic curves just for reference.}
\label{fig6}
\end{center}
\end{figure*}

We plot the overlap reduction functions in Fig.\,\ref{fig6} for existing detector pairs on the Earth, whose relative coordinates $(\beta,\;\sigma_{+},\;\sigma_{-})$ are listed in Table \ref{tab2}.

Let us begin with the explanation for the massless case. The overlap reduction functions start to oscillate and decay rapidly above the characteristic frequency given by $f_c \equiv c/(2 \pi |\Delta \mathbf{X}|)$. At low frequencies, the functions approach constant values, which are determined by the relative orientations of the detector pair. The difference between the polarization modes starts to appear at around the characteristic frequency. Mathematically, this is because in Eqs.\,(\ref{eq32}) - (\ref{eq37}) the coefficients of $j_0$ in the overlap reduction functions are exactly the same, while the coefficients of $j_2$ and $j_4$ are different. To this end, at the low frequency limit ($j_0\rightarrow1$, $j_2\rightarrow0$, and $j_4\rightarrow0$), all overlap reduction functions degenerate. 

On the other hand, in the massive case, overlap reduction functions have a cutoff at the frequency $f_g$, at which the phase velocity diverges and $\alpha$ is zero. At the intermediate frequencies $f>f_g$, the detector distance $|\Delta \vec{X}|$ in $\alpha$ is effectively reduced because $|\Delta \vec{X}|/c$ in the massless case is replaced with $|\Delta \vec{X}| /v_p$ in the massive case. That is, signal correlation between the detectors becomes stronger than the massless case at a fixed frequency. At high frequencies, as the phase velocity approaches the speed of light more and more, the overlap reduction functions finally coincide with
those in the massless case. As a result, the overlap reduction functions have frequency dependence as if they are shrunk in frequency, particularly at low frequencies. Interestingly, in the massive case, detectors are much more sensitive to a GWB at slightly higher frequencies than $f_g$, compared with the massless case.

\section{Sensitivity to graviton mass}
\label{sec4}

\subsection{Fisher matrix}
\label{sec:fisher-matrix}
To estimate measurement accuracy of graviton mass from the detection of a stochastic GWB, we use the Fisher information matrix derived in Appendix \ref{appA}:
\begin{align}
F_{ab}(\vec{\theta}) &= C_0 T \sum_{i=1}^{N_{\rm{pair}}} \int_{0}^{\infty} \frac{\Gamma_{ab,i} (f;\vec{\theta})}{{\cal{N}}_i (f) f^6} df \;, 
\label{eq56} \\
C_0 &\equiv 2 \left( \frac{3 H_0^2}{10 \pi^2} \right)^2 \;, \nonumber \\
\Gamma_{ab,i} (f; \vec{\theta}) &\equiv \sum_{A,A^{\prime}} \left[ \gamma_i^A \gamma_i^{A^{\prime}} (\partial_a \Omega_{\rm{gw}}^A) (\partial_b \Omega_{\rm{gw}}^{A^{\prime}}) \right. \nonumber \\
& \quad \quad \quad + (\partial_a \gamma_i^A) (\partial_b \gamma_i^{A^{\prime}}) \Omega_{\rm{gw}}^A \Omega_{\rm{gw}}^{A^{\prime}} \nonumber \\
& \quad \quad \quad + \gamma_i^A (\partial_b \gamma_i^{A^{\prime}}) \Omega_{\rm{gw}}^{A^{\prime}} (\partial_a \Omega_{\rm{gw}}^A) \nonumber \\
& \left. \quad \quad \quad +\gamma_i^{A^{\prime}}  (\partial_a \gamma_i^A) \Omega_{\rm{gw}}^A (\partial_b \Omega_{\rm{gw}}^{A^{\prime}}) \right] \;.
\label{eq55}
\end{align}
Here $\partial_a$ is the derivative with respect to $\theta_a$, and the product of the noise spectra of $I$- and $J$-th detectors is ${\cal{N}}_i(f) =P_I(f)P_J(f)$, which is defined for the $i=IJ$ detector pair. In our model of the GWB spectrum given in Eq.~(\ref{eq54}), each polarization is independent and we have in total 9 model parameters: $\vec{\theta}=\{ \Omega_{{\rm{gw}},0}^A,\,n_t^A,\,m_g^A \}$ for $A=T,V,S$. Note that the free parameters $m_g^A$ are included not only in $\Omega_{\rm{gw}}^A$ but also in $\gamma_i^A$ so that we have four terms in Eq.~(\ref{eq55}). Given the numerically evaluated Fisher matrix, the marginalized 1-$\sigma$ error of a parameter, $\Delta\theta_a$, is estimated from the inverse Fisher matrix 
\begin{equation}
\Delta \theta_a = \sqrt{\{\mathbf{F}^{-1}\}_{aa}}.
\end{equation}
It is well known that parameter estimation with the Fisher matrix is valid only when SNR is high enough, e.g. $\gtrsim10$ \cite{Cutler:2007mi,Vallisneri:2007ev}. For low SNR, the parameter estimation is too optimistic. Therefore, as explained later, we require at least ${\rm{SNR}}=10$ to estimate measurement accuracy of parameters.

By assuming that only a single polarization is present and explicitly assigning the raw and column of the Fisher matrix, the above expression of $F_{ab}$ is much simplified. In the presence of multiple polarization modes, adopting the method proposed in \cite{Nishizawa:2009bf,Nishizawa:2009jh}, we can separate the mixture of polarization signals and obtain the correlation signal with a single polarization mode. Thus in this section we assume that only a single polarization is present and omit the superscript $A$ and $A^{\prime}$ in the  expressions below in this section. The parameter estimation in the presence of multiple polarization modes will be discussed in the later section. Depending on whether the Fisher matrix contains the derivative with respect to $m_g$ or not, it is classified into three cases: 
\begin{description}
\item[(i)] $\theta_{a,b}=\{ \Omega_{{\rm{gw}},0},\,n_t \}$. 
\item[(ii)] $\theta_a=\{ \Omega_{{\rm{gw}},0},\,n_t \}$ and $\theta_b=m_g$.
\item[(iii)] $\theta_{a,b}=m_g$.
\end{description}

For $\theta_{a,b}=\{ \Omega_{{\rm{gw}},0},\,n_t \}$, since $\gamma_i$ contains only the parameter $m_g$, only the first term in Eq.~(\ref{eq55}) remains:
\begin{align}
F_{ab} &= C_0 T \sum_{i=1}^{N_{\rm{pair}}} \int_{0}^{\infty} \frac{\gamma_i^2 (\partial_a \Omega_{\rm{gw}}) (\partial_b \Omega_{\rm{gw}})}{{\cal{N}}_i (f) f^6} df \nonumber \\
& {\rm{for}} \;\; \theta_{a,b}=\{ \Omega_{{\rm{gw}},0},\,n_t \} \;.
\end{align}
For $\theta_a=\{ \Omega_{{\rm{gw}},0},\,n_t \}$ and $\theta_b=m_g$, the second and fourth terms in Eq.~(\ref{eq55}) vanish,
\begin{equation}
\Gamma_{am,i} (f; \vec{\theta}) = \gamma_i^2 (\partial_a \Omega_{\rm{gw}}) (\partial_m \Omega_{\rm{gw}}) + \gamma_i \Omega_{\rm{gw}} (\partial_m \gamma_i) (\partial_a \Omega_{\rm{gw}}) \;,
\end{equation}
where $\partial_m$ denotes the derivative with respect to $m_g$. Using
\begin{equation}
\partial_m \Omega_{\rm{gw}} (f) = -\frac{1}{2\pi} \Omega_{{\rm{gw}},0} \left( \frac{f}{f_0} \right)^{n_t} \delta [f-f_g] \;, 
\label{eq26}
\end{equation}
and performing the frequency integral in Eq.~(\ref{eq56}), we have
\begin{align}
F_{am} &= C_0 T \sum_{i=1}^{N_{\rm{pair}}} \left[ \int_{0}^{\infty} \frac{\gamma_i \Omega_{\rm{gw}} (\partial_m \gamma_i) (\partial_a \Omega_{{\rm{gw}}})}{{\cal{N}}_i f^6} df \right. \nonumber \\
& \left. -\frac{1}{2\pi} \left\{ \Omega_{{\rm{gw}},0} \left( \frac{f}{f_0} \right)^{n_t} \frac{\gamma_i^2 (\partial_a \Omega_{\rm{gw}})}{{\cal{N}}_i f^6}  \right\} \bigg|_{f=f_g} \right] \;, \nonumber \\
& {\rm{for}} \;\; \theta_{a}=\{ \Omega_{{\rm{gw}},0},\,n_t \} \;\; {\rm{and}} \;\; \theta_{b}=m_g \;.  
\end{align}
For $\theta_a, \theta_b=m_g$, substituting Eq.~(\ref{eq26}) for Eq.~(\ref{eq56}) and again performing the frequency integral give
\begin{align}
F_{mm} &= C_0 T \sum_{i=1}^{N_{\rm{pair}}} \left[ \int_{0}^{\infty} \frac{(\partial_m \gamma_i)^2 \Omega_{\rm{gw}}^2}{{\cal{N}}_i f^6} df \right. \nonumber \\
& -\frac{1}{\pi} \left\{ \Omega_{\rm{gw},0}^2 \left( \frac{f}{f_0} \right)^{2 n_t} \frac{\gamma_i (\partial_m \gamma_i)}{{\cal{N}}_i f^6} \right\} \bigg|_{f=f_g} \nonumber \\
& \left.+ \frac{1}{4\pi^2} \left\{ \Omega_{\rm{gw},0}^2 \left( \frac{f}{f_0} \right)^{2 n_t} \frac{\gamma_i^2}{{\cal{N}}_i f^6} \delta[f-f_g] \right\} \bigg|_{f=f_g} \right] \;.
\end{align}
In the third term of the above equation, $\delta (0)$ appears since the delta function has a support exactly at $f=f_g$ and otherwise zero. However in a real detector we cannot exactly measure the frequency $f_g$ and then we ignore the third term.    

To compute the Fisher matrix, what we need is the expressions for $\partial_a \Omega_{\rm{gw}}$ and $\partial_m \gamma_i$. From Eqs.~(\ref{eq54}), 
\begin{align}
\frac{\partial \Omega_{\rm{gw}}(f)}{\partial \Omega_{\rm{gw},0}} &= \left( \frac{f}{f_0} \right)^{n_t} \Theta [f-f_g] \;, \\
\frac{\partial \Omega_{\rm{gw}}(f)}{\partial n_t} &= \Omega_{\rm{gw}}(f) \log \left[ \frac{f}{f_0} \right] \;. 
\end{align}
As for $\partial_m \gamma_i$, first let us write the overlap reduction functions in Eqs.~(\ref{eq38}) - (\ref{eq37}) in a compact form as
\begin{align}
\gamma^A (\alpha,\, \beta,\,\sigma_{+},\,\sigma_{-} ) &= \Theta_{A+}(\alpha,\, \beta)\,\cos(4\sigma_{+}) \nonumber \\
&\;\; + \Theta_{A-}(\alpha,\, \beta)\,\cos(4\sigma_{-}) \;, 
\label{eq27} 
\end{align}
with
\begin{align}
\Theta_{A+}(\alpha,\, \beta) &= a_0^A(\beta) j_0 + a_2^A(\beta) j_2 + a_4^A(\beta) j_4 \;, \nonumber \\
\Theta_{A-}(\alpha,\, \beta) &= b_0^A(\beta) j_0 + b_2^A(\beta) j_2 + b_4^A(\beta) j_4 \;. \nonumber
\end{align}
Differentiating Eq.~(\ref{eq27}) with respect to $m_g^A$ and using the recursion relation of the spherical Bessel functions, we have 
\begin{align}
\partial_m \gamma^A &= -\frac{\sqrt{v_p^2-1}}{\omega_c} \left[ (\partial_\alpha \Theta_{A+}) \,\cos(4\sigma_{+}) \right. \nonumber \\
& \quad \quad \quad \quad \quad \quad \left.+ (\partial_\alpha \Theta_{A-})\,\cos(4\sigma_{-}) \right] \;, \nonumber \\
\partial_\alpha \Theta_{A+} &= \left(-a_0+\frac{2}{5}a_2 \right) j_1 + \left(-\frac{3}{5} a_2+\frac{4}{9}a_4 \right) j_3 -\frac{5}{9}a_4 j_5 \;, \\
\partial_\alpha \Theta_{A-} &= \left(-b_0+\frac{2}{5}b_2 \right) j_1 + \left(-\frac{3}{5} b_2+\frac{4}{9}b_4 \right) j_3 -\frac{5}{9}b_4 j_5 \;, 
\end{align}
where the coefficients $a_i$, $b_i$ ($i=0,2,4$) are functions of $\beta$ and different for each polarization mode, and $\omega_c \equiv c/|\Delta \mathbf{X}|$.

\subsection{Signal to noise ratio}

\begin{figure*}[t]
\begin{center}
\includegraphics[width=15cm]{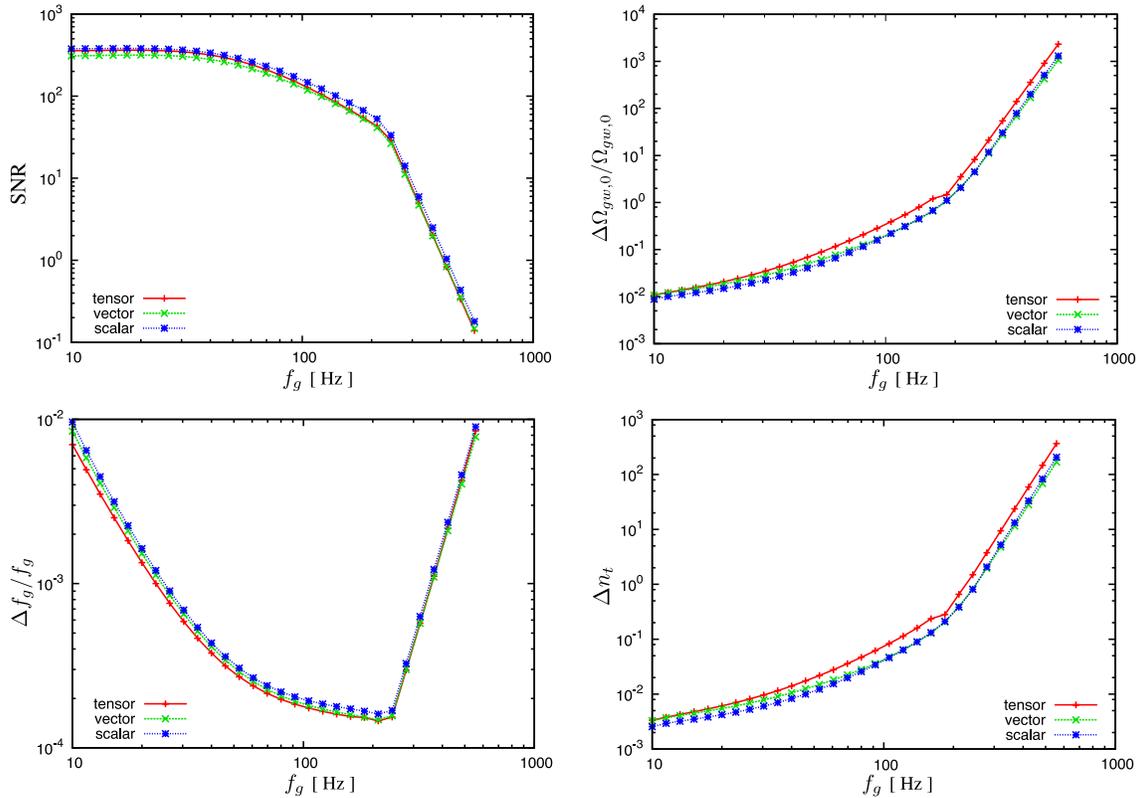}
\caption{SNR and the parameter estimation accuracy as a function of a fiducial value of $f_g$ in Hz in the presence of a single polarization mode. The fiducial values of the other parameters are $h_0^2 \Omega_{{\rm{gw}},0}=10^{-7}$ and $n_t=0$. The top-left is SNR, the top-right is $\Delta \Omega_{{\rm{gw}},0}/\Omega_{{\rm{gw}},0}$, the bottom-left is $\Delta f_g/f_g$, the bottom-right is $\Delta n_t$. The red, green, and blue curves are tensor, vector, and scalar modes, respectively.}
\label{fig:single-net}
\end{center}
\end{figure*}

The parameter estimation with the Fisher information is valid in a high SNR regime. If the SNR is low, e.g. ${\rm{SNR}}< 10$, the measurement accuracy of the parameters could be overestimated. To make the results derived with the Fisher matrix reliable, we will first check SNRs for a stochastic GWB.

Using the SNR formula in Eq.~(\ref{eq18}), we evaluate the SNR for each polarization mode with a detector pair, assuming that only one polarization mode (tensor, vector, or scalar mode) exists and 3-yr observation time. As for the power spectra of the detector noise $P_I(f)$, we use for simplicity that of aLIGO for all advanced detectors (K, H, L, V), which is based on \cite{LIGO:1999} and given by \cite{Seto:2008sr}
\begin{widetext}
\begin{equation}
P(f) = \left\{
\begin{array}{lll} 
\displaystyle
10^{-44} \left( \frac{f}{10\,{\rm Hz}} \right)^{-4} + 10^{-47.25} \left( \frac{f}{100\,{\rm Hz}} \right)^{-1.7} {\rm Hz}^{-1} 
\quad \quad \quad \quad \;\;\; {\rm{for}}\;\; 10\,{\rm Hz} \leq f \leq 240\,{\rm Hz} \;,  \nonumber \\
\displaystyle
10^{-46} \left( \frac{f}{1000\,{\rm Hz}} \right)^{3} {\rm Hz}^{-1} 
\quad \quad \quad \quad \quad \quad \quad \quad \quad \quad \quad \quad \quad \quad \quad \;\; {\rm{for}}\;\; 240\,{\rm Hz} \leq f \leq 3000\,{\rm Hz} \;, \nonumber \\
\displaystyle 
\infty \quad \quad \quad \quad \quad \quad \quad \quad \quad \quad \quad \quad \quad \quad \quad \quad \quad \quad \quad \quad \quad \quad \quad \quad \quad \; {\rm otherwise} \;.
\end{array}
\right. 
\label{eq58}
\end{equation}
\end{widetext}
The lower and higher cutoffs of frequency integral is set to $f_{\rm{low}}=10\,{\rm{Hz}}$ and $f_{\rm{high}}=1\,{\rm{kHz}}$. 

For a detector network, $({\rm{SNR}})^2$ is given by the squared sum of SNR of each detector pair,
\begin{equation}
({\rm{SNR}})^2=\sum_{i=1}^{N_{\rm{pair}}} ({\rm{SNR}}_i)^2\;, \nonumber
\end{equation}
where $N_{\rm{pair}}$ is the number of detector pairs. In Fig.~\ref{fig:single-net}, we show the SNR of a detector network (K, H, L, V) for a stochastic GWB of $h_0^2 \Omega_{{\rm{gw}},0}^A=10^{-7}$ and $n_t^A=0$ as a function of graviton-mass frequency $f_g^A$. At low $f_g$, since the GWB spectrum is indistinguishable from the massless case, the SNR coincides with that in the massless case. As $f_g$ increases, the SNR gradually decreases due to the low frequency cutoff at $f=f_g$ of the GWB spectrum. For the fiducial parameters $h_0^2 \Omega_{{\rm{gw}},0}^A=10^{-7}$ and $n_t^A=0$, if we require ${\rm{SNR}}=10$ for reliable detection, this implies that massive GWB with $f_g$ less than $\sim 300\,{\rm{Hz}}$ is detectable. As is clear from the SNR formula in Eq.~(\ref{eq18}), the SNR scales linearly with $\Omega_{{\rm{gw}},0}$. For $h_0^2 \Omega_{{\rm{gw}},0}^A=10^{-8}$, the SNR is degraded but is detectable if $f_g \lesssim 150\,{\rm{Hz}}$.

\subsection{Sensitivity to graviton mass}

We calculate the Fisher matrix with free parameters, $\{ \Omega_{{\rm{gw}},0}^A,\,n_t^A,\,m_g^A \}$, $A=T, V, S$, assuming that only a single polarization mode exists (The superscript $A$ is omitted below). The parameter estimation accuracy with a detector network is computed by adding the Fisher matrices of all detector pair. The pivot frequency $f_0$ of the spectral index $n_t$ is chosen as $f_0=1\,{\rm{Hz}}$, which is outside of the observation frequency band in order to avoid insensitivity to $n_t$ at $f=f_0$. The fiducial parameters are set to $h_0^2 \Omega_{{\rm{gw}},0}=10^{-7}$ and $n_t=0$. 

The results are shown in Fig.~\ref{fig:single-net} as a function of the fiducial values of $f_g$. The measurement errors of $\Omega_{{\rm{gw}},0}$ and $n_t$ are smaller at low $f_g$ and approach those in the massless case at the limit of $f_g\rightarrow 0$. This is because the contribution to the Fisher matrix is an integrated quantity in frequency. As $f_g$ increases, the measurement accuracies of $\Omega_{{\rm{gw}},0}$ and $n_t$ are deteriorated. At $f_g$ slightly below the high frequency cutoff required by the SNR threshold 10, i.e. $\sim 200\,{\rm{Hz}}$, the parameters $\Omega_{{\rm{gw}},0}$ and $n_t$ are still measurable. On the other hand, for the measurement accuracy of $m_g$ or $f_g$, the detector network is most sensitive to the value of $f_g$ around the minimum of the detector noise curve, $\sim 100\,{\rm{Hz}}$. This means that the dominant contribution to the Fisher matrix comes from the frequencies around $f\sim f_g$. Interestingly, the parameter $f_g$ is well determined for all $f_g$ in the frequency band of a ground-based detector.      

\begin{figure}[t]
\begin{center}
\includegraphics[width=8cm]{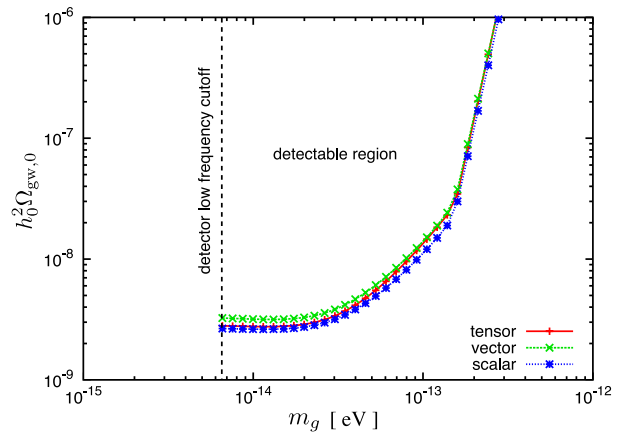}
\caption{Detectable parameter region of $m_g$ and $\Omega_{{\rm{gw}},0}$ when $n_t^A=0$. The red, green, and blue curves are detection threshold for tensor, vector, and scalar modes, respectively. The region above the curves is detectable with a detector network.}
\label{fig:mg-OmegaGW0}
\end{center}
\end{figure}

From the above results, we conclude that graviton mass is detectable for a massive GWB of $h_0^2 \Omega_{{\rm{gw}},0}^A=10^{-7}$ and $n_t^A=0$ if the graviton mass is in the range
\begin{align}
&6.6\times10^{-15}\,{\rm{eV}} \leq m_g^A \leq 1.8\times10^{-13}\,{\rm{eV}}\;, \nonumber \\
& \quad \quad \quad \quad \quad \quad {\rm{for}} \;\; A=T, V, S 
\end{align}
When the amplitude of a GWB is different from $h_0^2 \Omega_{{\rm{gw}},0}^A=10^{-7}$, the Fisher matrix just scales with the square of the amplitude. Thus the measurement accuracy of $f_g$ inversely scales with the amplitude, namely, $\Delta f_g/f_g \propto (\Omega_{{\rm{gw}},0})^{-1}$. We also have another criterion for mass detection, ${\rm{SNR}}>10$, which guarantees reliable parameter estimation based the Fisher matrix. As seen in Fig.~\ref{fig:single-net}, the SNR criterion is tighter in our calculation. Then we obtain detectable parameter region of $m_g$ and $\Omega_{{\rm{gw}},0}$ when $n_t^A=0$, which is shown in Fig.~\ref{fig:mg-OmegaGW0}. Below the curves, GWB is not detected with sufficient SNR and graviton mass cannot be determined. The slight differences of the detection threshold of graviton mass are due to the small differences of the antenna pattern function of each polarization mode.

One might wonder how the detectable parameter region of graviton mass is changed if the fiducial value of $n_t$ deviates from zero. Since our choice of the pivot frequency of the GWB spectrum is $f_0=1\,{\rm{Hz}}$, blue (red) tilt of the spectrum just increases (decreases) SNR more at high frequencies. Then positive (negative) $n_t$ expands (reduces) the detectable parameter region of graviton mass in Fig.~\ref{fig:mg-OmegaGW0}.

Finally we remark that we can set an upper limit on graviton mass if a GWB is detected but the lower frequency cutoff is not detected. In this case, the lower frequency cutoff of the GWB spectrum is at least below the frequency band of a detector. Therefore we find the mass limit $f_g<f_{\rm{low}}$, where $f_{\rm{low}}$ is the lower cutoff of detector noise curve and say, $\sim 10\,{\rm{Hz}}$. In this case, the mass limit will be $m_g<6.6\times10^{-15}\,{\rm{eV}}$. For the tensor mode, this mass constraint is much weaker than what obtained from the observation of binary pulsars, $m_g < 7.6\times 10^{-20}\,{\rm{eV}}$ \cite{Finn:2001qi}. But it is worthy that the bounds on graviton mass are obtained from different observations of GWs at completely different environments at binary pulsars and on the Earth, because it is possible to give gravitons larger mass on the Earth if an environment-dependent screening effect such as the chameleon mechanism would occur.

\section{Mixture of polarization modes and its separability}
\label{sec5}
In the previous section, we computed the parameter estimation accuracy of the GWB spectrum, particularly focusing on the measurement accuracy of the graviton mass, in the presence of a single polarization mode. However, if there exist massive polarization modes due to the modification of gravity theory, what we observe is likely to be a mixture signal of different polarizations. For example, if a gravity theory has an additional scalar degree of freedom, a GWB would be composed of ordinary massless tensor mode and massive scalar mode. To this end, in this section we investigate the parameter estimation in the presence of a mixed GWB. We assume that a tensor graviton is massless, $m_g^T=0$ \footnote{In the actual calculation, we set $f_g=10^{-2}\,{\rm{Hz}}$ just for convenience of numerical computation.}, since graviton mass in a tensor mode has already been tightly constrained as mentioned in Sec.~\ref{sec:graviton-mass-constraints}. Also for simplicity we consider the case of flat GWB spectra, $n_t^A=0$, $A=T,V,S$. With these fiducial values, we estimate the measurement accuracy of five parameters: $\Omega_{{\rm{gw}},0}^A$ ($A=T,V,S$), $m_g^V$, $m_g^S$.

To do so, we need extra terms of the Fisher matrix arising from cross correlation between different polarization modes, in addition to the auto-correlation terms provided in Sec.~\ref{sec:fisher-matrix}. The explicit forms of the elements of the Fisher matrix are obtained in the same way as in Sec.~\ref{sec:fisher-matrix} and are given in Appendix \ref{appB}.  

\begin{figure}[h]
\begin{center}
\includegraphics[width=8cm]{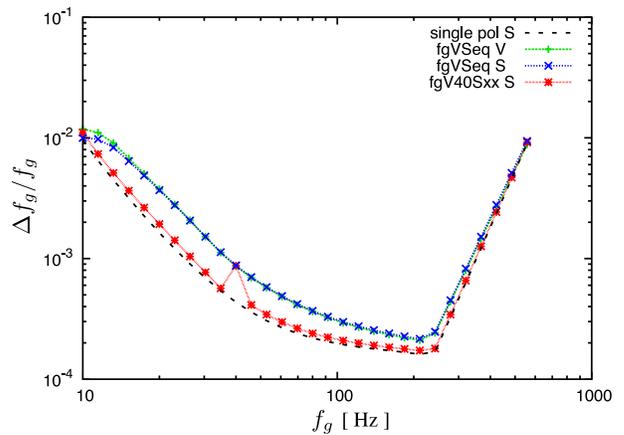}
\caption{Measurement accuracy of graviton mass when all polarization signals are mixed and each mode has different graviton mass. In the legend, "fgVSeq V" and "fgVSeq S" are the measurement accuracies of $f_g^V$ and $f_g^S$ when both have equal masses $f_g=f_g^V=f_g^S$. While "fgV40Sxx S" is the measurement accuracy of $f_g^S$ when the vector graviton mass is fixed to $f_g^V=40\,{\rm{Hz}}$ and $f_g=f_g^S$. The dashed curve is the one we obtained in the previous section in the presence of a scalar polarization mode. In all cases, the fiducial amplitude is $h_0^2 \Omega_{{\rm{gw}},0}^A=10^{-7}$ for all polarization modes.}
\label{fig:mix-net-mg}
\end{center}
\end{figure}

First, we study the effect of mass degeneracy. The fiducial values of the spectral amplitude are set to $h_0^2 \Omega_{{\rm{gw}},0}^A=10^{-7}$ for all polarization modes. The vector and scalar graviton masses are set to different values. The result is shown in Fig.~\ref{fig:mix-net-mg}. When the vector and scalar graviton masses are equal, there is weak degeneracy of the parameters, which slightly worsens the measurement accuracies of $f_g^V$ and $f_g^S$. If the vector graviton mass is set to $f_g^V=40\,{\rm{Hz}}$, the measurement accuracy of $f_g^S$ is improved because of the broken degeneracy of the parameters, and is almost the same as the one we obtained in the previous section in the presence of a single scalar polarization mode. This indicates that in general the mixture of polarization signals can be well separated by data-analysis procedures unless the parameters are degenerated. Such a method has already proposed and investigated in \cite{Nishizawa:2009bf,Nishizawa:2009jh} in the case of the massless GWB. The authors have shown that at least three correlation signals allow one to separate three polarization modes, without any degeneracy between polarizations in the case of advanced ground-based detectors that we consider in this paper. So the good separability of the polarization modes is manifest even if graviton has its mass. 

\begin{figure}[h]
\begin{center}
\includegraphics[width=8cm]{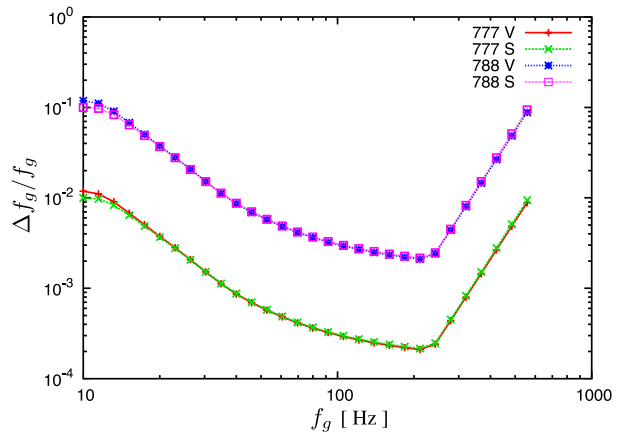}
\caption{Measurement accuracy of graviton mass when all polarization signals are mixed and each mode has different GWB amplitude. Vector and scalar graviton masses are set to $f_g=f_g^V=f_g^S$. In the legend, "777 V" and "777 S" represent the measurement accuracies of $f_g^V$ and $f_g^S$ when the fiducial amplitude is $h_0^2 \Omega_{{\rm{gw}},0}^A=10^{-7}$ for all polarization modes.  "788 V" and "788 S" are those when the fiducial amplitudes are $h_0^2 \Omega_{{\rm{gw}},0}^T=10^{-7}$ and $h_0^2 \Omega_{{\rm{gw}},0}^V=h_0^2 \Omega_{{\rm{gw}},0}^S=10^{-8}$.}
\label{fig:mix-net-OmegaGW0}
\end{center}
\end{figure}

Second, we investigate how different amplitude between the polarization modes affects the measurement accuracy of graviton mass. In Fig.~\ref{fig:mix-net-OmegaGW0}, the result is shown. If a GWB spectrum is the sum of large tensor mode with $h_0^2 \Omega_{{\rm{gw}},0}^T=10^{-7}$ and small vector and scalar modes with $h_0^2 \Omega_{{\rm{gw}},0}^{V,S}=10^{-8}$, the measurement accuracies of $f_g^V$ and $f_g^S$ just inversely scales with the power amplitude. Again we can conclude that the mixture of polarization signals can be well separated by data-analysis procedure even if the GWB spectrum of each polarization mode has different amplitude.

\section{Conclusions}
\label{sec6}

We have studied the search method for a massive stochastic GWB, extending the ordinary method for massless graviton to that for massive graviton and allowing tensor, vector, and scalar polarization modes of the GWB. If a GWB is massive, the phase velocity, which is larger than the speed of light, affects the cross-correlation statistics by changing effective distance between detectors. Also there is a lower frequency cutoff on a GWB spectrum that corresponds to graviton mass. 

To investigate the detectability of graviton mass with a ground-based detectors, we used a generically parameterized model of a GWB spectrum and estimated the measurement accuracies of the parameters with the Fisher information matrix. As a results, we found that if a GWB is detected at the level of $h_0^2 \Omega_{{\rm{gw}},0}=10^{-7}$, we can determine the mass of graviton in the range of $7\times 10^{-15}\,{\rm{eV}} \lesssim m_g \lesssim 2\times 10^{-13}\,{\rm{eV}}$ for each polarization mode with a ground-based detectors. We also showed that even if the GWB signal is a mixture of three polarization modes, we can safely separate them and determine the mass of graviton. Even if a GWB is detected but the lower frequency cutoff is not detected, we can set an upper limit on graviton mass, which is the lower cutoff of detector noise curve, say, $\sim 10\,{\rm{Hz}}$ or $\sim 7\times 10^{-15}\,{\rm{eV}}$.

Finally note that our method to search for graviton mass can be a quite generic and model-independent test of alternative theory of gravity, because we do not assume any specific model of gravity theory in the reconstruction of a GWB spectral shape. The constraints on the polarization degrees of freedom, graviton mass, spectral index, and spectral amplitude would help select out or exclude some specific models of gravity. In this paper, although we focused on ground-based detectors such as aLIGO, aVIRGO, and KAGRA, it would be interesting to consider the constraints obtained in the future with other detectors such as ET \cite{ETdesign}, LISA \cite{AmaroSeoane:2012je}, DECIGO \cite{Kawamura:2011zz}, BBO \cite{Harry:2006fi}, and pulsar timing array \cite{Joshi:2013at}, since different observational frequency bands correspond to different mass ranges of graviton. In particular, among GW detectors, LISA is sensitive in lowest frequency band as low as $\sim10^{-4}\,{\rm{Hz}}$, which is roughly 5 - 6 orders lower than the frequency band of a ground-base detector. If a GWB is detected by LISA but the low frequency cutoff of the spectrum is not detected, the graviton mass constraint for the tensor mode would be comparable to the binary pulsar constraint less than $\sim10^{-20}\,{\rm{eV}}$.

\begin{acknowledgments}
We are much obliged to an anonymous referee for his/her valuable comments to improve the manuscript. A.N. would like to thank N. Seto for helpful discussions. K.H. would like to thank to Bruce Allen for warm hospitality during his stay in Hannover and Soumya D. Mohanty for valuable comments and encouragement. A.N. and K.H. are supported by Grant-in-Aid for Scientific Research on Innovative Areas, No.24103006.
\end{acknowledgments}

\appendix

\section{Derivation of Fisher matrix}
\label{appA}
Here we will derive the Fisher matrix for estimating measurement accuracy of the parameters of a GWB. When one correlates detector signals in a frequency bin $\Delta f$, the estimated value of the correlation signal of $i$-th detector pair (for I-th and J-th detector pair, $i=IJ$), $\hat{\mu}_i(f)$, fluctuates around the true value, $\mu_i(f;\vec{\theta})$, where $\vec{\theta}$ is a set of estimated parameters. Assuming the width of a single frequency bin that we analyze is much larger than the frequency resolution of the data but enough smaller so as to neglect the frequency dependence of physical quantity, a likelihood function for $\hat{\mu}_i(f)$ is expected to be Gaussian distribution, owing to the central limit theorem. Then the multidimensional likelihood function for $N_{\rm{pair}}$ detector pairs is written as
\begin{equation}
L\bigl[ \{ \hat{\mu}_i(f) \} \bigr] \propto \exp \left[ -\sum_{i=1}^{N_{\rm{pair}}} \frac{\{ \hat{\mu}_i(f)-\mu_i(f;\vec{\theta}) \}^2}{2 \sigma_i^2(f)} \right] \;.
\label{eq51}
\end{equation}
Suppose that $\vec{\theta}_{ML}$ is the maximum likelihood values of the parameters and denote the small deviation from it by $\Delta \vec{\theta}$. The likelihood function can be expanded around its peak $\vec{\theta}_{ML}$ and by neglecting the first derivative it reduced to 
\begin{equation}
L\bigl[ \{ \hat{\mu}_i(f) \} \bigr] \propto \exp \left[ -\sum_{i=1}^{N_{\rm{pair}}} \sum_{a,b} \frac{\partial_a \mu_i(f;\vec{\theta}) \partial_b \mu_i(f;\vec{\theta})}{2 \sigma_i^2(f)} \right] \;.
\label{eq52}
\end{equation} 
From this expression, the Fisher matrix reads 
\begin{equation}
F_{ab}(f) = \sum_{i=1}^{N_{\rm{pair}}} \frac{\partial_a \mu_i(f;\vec{\theta}) \partial_b \mu_i(f;\vec{\theta})}{2 \sigma_i^2(f)}\;.
\label{eq53}
\end{equation}
The correlation signal and noise variation are given in Eqs.~(\ref{eq30}) and (\ref{eq15})
as 
\begin{align}
\mu_i (f;\vec{\theta}) &= \frac{3H_0^2}{10\pi^2} T \, f^{-3} \sum_A \gamma_i^A (f;\vec{\theta}) \Omega_{\rm{gw}}^A (f;\vec{\theta}) \Delta f \;, \label{eq30a} \\
\sigma ^2 (f) &= \frac{T}{2} \,{\cal{N}}_i (f)\, \Delta f \:,
\label{eq15a}
\end{align}
where we defined ${\cal{N}}_i (f) \equiv P_I(f) P_J(f)$. Note that these quantities are defined in a positive frequency range, $0<f<\infty$, and a factor of two differs from Eqs.~(\ref{eq30}) and (\ref{eq15}). Substituting Eqs.~(\ref{eq30a}) and (\ref{eq15a}) into Eq.~(\ref{eq53}) and integrating over frequency, we obtain
\begin{align}
F_{ab} &= \sum_{i=1}^{N_{\rm{pair}}} 2 \left( \frac{3 H_0^2}{10 \pi^2} \right)^2 T \int_{0}^{\infty} \frac{\Gamma_{ab} (f;\vec{\theta})}{{\cal{N}}_i (f) f^6} df \;, \\
\Gamma_{ab} (f; \vec{\theta}) &\equiv \sum_{A,A^{\prime}} \left[ \gamma_i^A \gamma_i^{A^{\prime}} (\partial_a \Omega_{\rm{gw}}^A) (\partial_b \Omega_{\rm{gw}}^{A^{\prime}}) \right. \nonumber \\
& \quad \quad \quad + (\partial_a \gamma_i^A) (\partial_b \gamma_i^{A^{\prime}}) \Omega_{\rm{gw}}^A \Omega_{\rm{gw}}^{A^{\prime}} \nonumber \\
& \quad \quad \quad + \gamma_i^A (\partial_b \gamma_i^{A^{\prime}}) \Omega_{\rm{gw}}^{A^{\prime}} (\partial_a \Omega_{\rm{gw}}^A) \nonumber \\
& \quad \quad \quad \left. +\gamma_i^{A^{\prime}}  (\partial_a \gamma_i^A) \Omega_{\rm{gw}}^A (\partial_b \Omega_{\rm{gw}}^{A^{\prime}}) \right] \;.
\end{align}
Here $\partial_a$ is the derivative with respect to $\theta_a$. 

\section{Cross-correlated terms of Fisher matrix between different polarizations}
\label{appB}
To calculate the measurement accuracy of graviton mass for a mixture signal of GWBs in different polarization modes, we need extra terms of the Fisher matrix arising from cross correlation between different polarizations, in addition to the auto-correlation terms provided in Sec.~\ref{sec:fisher-matrix}. Here we write down an explicit forms of the cross-correlated terms between different polarizations. The extra terms are classified into three cases: 
\begin{description}
\item[(i)] $\theta_a=\Omega_{{\rm{gw}},0}^A,\;\; \theta_b=\Omega_{{\rm{gw}},0}^{A^{\prime}}, \;\; A \neq A^{\prime}$. 
\item[(ii)] $\theta_a=\Omega_{{\rm{gw}},0}^A,\;\; \theta_b=m_g^{A^{\prime}}, \;\; A \neq A^{\prime}$.
\item[(iii)] $\theta_a=m_g^A,\;\; \theta_b=m_g^{A^{\prime}}, \;\; A \neq A^{\prime}$.
\end{description}

The case (i) has the additional term
\begin{align}
\Gamma_{\Omega \Omega,i}^{AA^{\prime}} (f; \vec{\theta}) &=\gamma_i^A \gamma_i^{A^{\prime}} (\partial_a \Omega_{\rm{gw}}^A) (\partial_b \Omega_{\rm{gw}}^{A^{\prime}}) \nonumber \\
&=\gamma_i^A \gamma_i^{A^{\prime}} \Theta \left[ f-f_g^A \right] \Theta \left[ f-f_g^{A^{\prime}} \right] \nonumber \\
&=\gamma_i^A \gamma_i^{A^{\prime}} \Theta \left[ f-{\rm{max}}(f_g^A,f_g^{A^{\prime}}) \right]  \;.
\end{align}
The frequency integral in Eq.~(\ref{eq56}) is performed above the frequency ${\rm{max}}(f_g^A,f_g^{A^{\prime}})$. We have
\begin{equation}
F_{\Omega \Omega}^{AA^{\prime}} = C_0 T \sum_{i=1}^{N_{\rm{pair}}} \int_{{\rm{max}}(f_g^A,f_g^{A^{\prime}})}^{\infty} \frac{\gamma_i^A \gamma_i^{A^{\prime}}}{{\cal{N}}_i (f) f^6} df \;.
\end{equation}

The case (ii) has the additional term
\begin{align}
\Gamma_{\Omega m,i}^{AA^{\prime}} (f; \vec{\theta}) &= -\frac{1}{2\pi} \Omega_{{\rm{gw}},0}^{A^{\prime}} \gamma_i^A \gamma_i^{A^{\prime}} \Theta \left[ f-f_g^A \right] \delta \left[ f-f_g^{A^{\prime}} \right]  \nonumber \\
&+ \gamma_i^A \Omega_{{\rm{gw}},0}^{A^{\prime}} (\partial_m \gamma_i^{A^{\prime}}) \Theta \left[ f-f_g^A \right] \;.
\end{align}
After the frequency integration, we have
\begin{align}
F_{\Omega m}^{AA^{\prime}} &= C_0 T \sum_{i=1}^{N_{\rm{pair}}} \left[ \int_{f_g^A}^{\infty} \frac{\gamma_i^A \Omega_{\rm{gw}}^{A^{\prime}} (\partial_m \gamma_i^{A^{\prime}})}{{\cal{N}}_i f^6} df \right. \nonumber \\
&- \left. \frac{\Omega_{{\rm{gw}},0}^{A^{\prime}}}{2\pi} \left. \left\{ \frac{\gamma_i^A \gamma_i^{A^{\prime}}}{{\cal{N}}_i f^6} \right\} \right|_{f=f_g^{A^{\prime}}} \Theta \left[ f_g^{A^{\prime}}-f_g^A \right]   \right] \;.
\end{align}

The case (iii) has the additional term
\begin{align}
F_{m m}^{AA^{\prime}} &= C_0 T\, \Omega_{\rm{gw}}^{A} \Omega_{\rm{gw}}^{A^{\prime}} \sum_{i=1}^{N_{\rm{pair}}} \nonumber \\
&\left[ \int_{{\rm{max}}(f_g^A,f_g^{A^{\prime}})}^{\infty} \frac{(\partial_m \gamma_i^A) (\partial_m \gamma_i^{A^{\prime}})}{{\cal{N}}_i f^6} df \right. \nonumber \\
&- \frac{1}{2\pi} \left. \left\{ \frac{\gamma_i^A (\partial_m \gamma_i^{A^{\prime}})}{{\cal{N}}_i f^6} \right\} \right|_{f=f_g^A} \Theta \left[ f_g^A-f_g^{A^{\prime}} \right] \nonumber \\
&- \left. \frac{1}{2\pi} \left. \left\{ \frac{ (\partial_m \gamma_i^A) \gamma_i^{A^{\prime}}}{{\cal{N}}_i f^6} \right\} \right|_{f=f_g^{A^{\prime}}} \Theta \left[ f_g^{A^{\prime}}-f_g^A \right] \right] \;.
\end{align}

\bibliography{/Volumes/USB-MEMORY/my-research/bibliography}

\end{document}